# A review of recent studies on nonlinear dynamics of microtubules and DNA[*]

**Slobodan Zdravković**[†]
Vinča Institute of Nuclear Sciences, University of Belgrade, P.O. Box 522, Atomic Physics Laboratory (040), 11001 Beograd, Serbia

ABSTRACT

Nonlinear dynamics of two biomolecules is studied. These are a microtubule and DNA molecule. Two mathematical procedures are explained, yielding to three kinds of solitary waves moving through the systems. These waves are kinks, modulated solitary waves called breathers and bell-type solitons.

## 1. Introduction

Biomolecules are nonlinear systems due to inevitable presence of weak chemical bonds. Namely, strong interactions yield to small amplitudes and we can assume that intensities of attractive and repulsive forces are equal. This means that these interactions can be modelled by harmonic potential energies, i.e. by functions of the type $f(x) = kx^2$, where $k = $ const. Its first derivative is a force, obviously a linear function, which brings about a linear differential equation (DE). As for weak interactions, they should be modelled by enharmonic potential energies, which yield to nonlinear DEs.

Two examples are studied in this article. These are microtubule (MT) and DNA. The structure of DNA is known. It consists of two mutually interacting strands. Each strand is a series of covalently interacting nucleotides. A covalent bond is the strongest chemical interaction, which means that the strands are linear structures.

[*] This work has been supported by funds from Project within the Cooperation Agreement between the JINR, Dubna, Russian Federation and Ministry of Education, Sciences and Technological Development of Republic of Serbia: Theory of Condensed Matter Physics.
[†] e-mail address: szdjidji@vin.bg.ac.rs

The strands are connected by weak hydrogen interactions and, of course, DNA as a whole is a nonlinear system.

MT is a long hollow cylindrical polymer structure that spreads between a nucleus and cell membrane [1,2]. Its surface is usually formed out of 13 long structures called protofilaments (PFs), as shown in Fig. 1. Each PF represents a series of electric dipoles called dimers, whose mass and length are $m = 1.8 \times 10^{-22}$ kg and $l = 8$nm, respectively [3,4]. The lengths of MTs vary from a few hundred nanometers up to meters in long nerve axons [5]. The longitudinal, tangential and radial components of electric dipole moment are: $p_z = 337$Debye, $p_\theta = 198$Debye and $p_r = -1669$Debye, respectively [6]. Hence, $p_z$ is in the direction of MT.

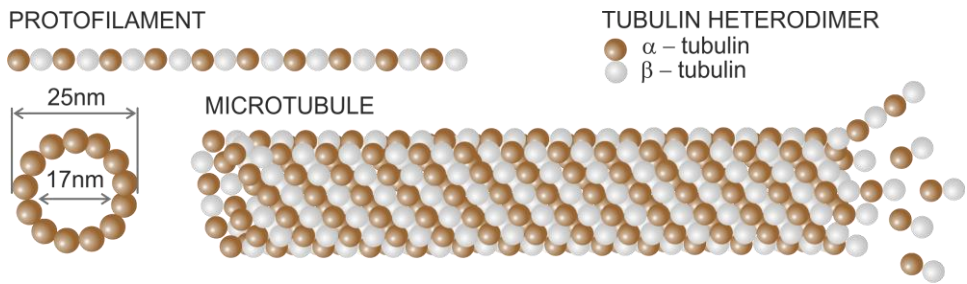

**Fig. 1.** Microtubule

A head-to-tail binding of dimers, resulting in PFs, appear to be much stronger than those between adjacent PFs [7,8]. This means that a single PF can be seen as a linear system and modelled by the harmonic potential energy. On the other hand, the interaction with the remaining dimers is usually modelled by the function of the type $F(x) = -ax^2 + bx^4 - cx$, where $a > 0$, $b > 0$ and $c \geq 0$ are assumed. For $c = 0$, the function $F(x)$ is symmetric, corresponding to so-called W-potential energy. Therefore, MT as a whole is a nonlinear system.

It was pointed out that the two biological systems are studied in this paper. To model them, two mathematical procedures will be used. They are semi-discrete approximation (SDA) and continuum approximation (CA). The combinations MT-SDA and DNA-CA are explained in Sections 2 and 3, respectively. In Section 4, we deal with DNA-RNA transcription, where SDA is used, while Section 5 is devoted to concluding remarks. It is interesting that the final results, i.e. the solutions of the mentioned DEs, depend on the used mathematical method rather than on the studied system. The SDA yields to modulated solitary waves called breathers, while the common solution corresponding to the CA is a kink soliton, or kink for short.

A general procedure is equal for both systems and both mathematical procedures. The first step is Hamiltonian. This is nothing but a collection of energies, describing existing interactions. We use generalized coordinates and well-known Hamilton`s equations to obtain dynamical equation of motion. The terms in this equation are forces. Finally, we solve this equation using aforementioned mathematical approximations.

## 2. Tangential model of microtubules and semi-discrete approximation

There are a couple of models describing nonlinear MT dynamics. Depending on a coordinate which determines a dimer`s displacement they can be either longitudinal or angular. Of course, two component models are also possible. We can monitor a certain evolution of the models through papers [9-17]. Ref. [9] describes the first model, a longitudinal one, where W-potential energy was introduced. Its improved version, which we call $u$-model, was described in Refs. [10-12]. It was shown that Morse potential may be used instead of the W-one [13]. An angular so-called $\varphi$-model, that does not comprise the W-potential energy, was introduced in Refs. [14] and [15], while its improved version, including this term, is called a general model [16].

This section is based on a recently introduced two component model that we call a tangential model (TM) [17]. The first that should be clarified is the W-potential energy, modelled by the function $F(x)$, as explained above. This function obviously has two minima, which means that there are two directions of electric field around which the dimer can oscillate. Let the appropriate electric field strengths be $\vec{E}_1$ and $\vec{E}_2$. A resultant internal electric field $\vec{E} = \vec{E}_1 + \vec{E}_2$, coming from all dimers, is in the direction of MT. In principle, the dimer can oscillate around this direction, but any displacement would move it towards the directions of either $\vec{E}_1$ or $\vec{E}_2$. This means that the dimer's position in the direction of $\vec{E}$ is not stable and corresponds to the maximum of the W-potential energy.

Fig. 2 shows oscillation of the dimer around $\vec{E}_1$ only. The orientation of this field is determined by $\theta_{01} \equiv \theta_0$, which is the angle between the direction of the PF and $\vec{E}_1$. A coordinate determining a displacement from the direction of $\vec{E}_1$ is $\varphi$, while the dimer's position with respect to the direction of the PF is $\theta$. It is obvious that

$$\theta = \theta_0 + \varphi \qquad (1)$$

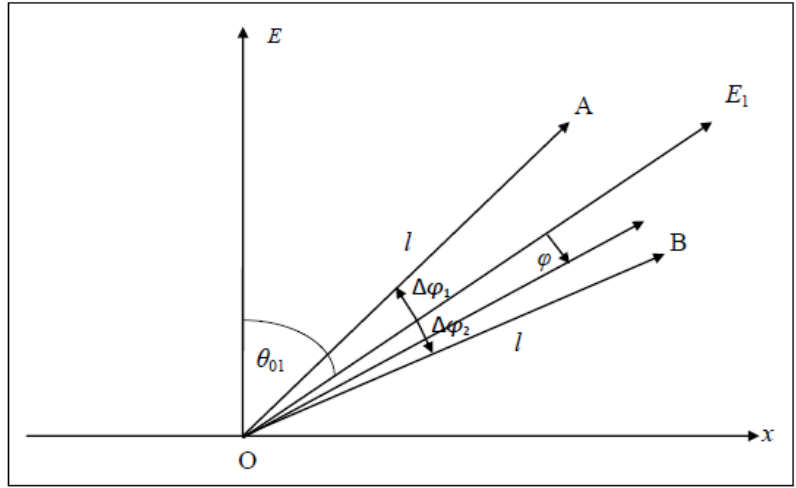

**Fig. 2.** A schematic representation of the dimer's oscillation

The three components of the electric dipole moment of the single dimer were defined above. They are in the direction of the MT ($\vec{p}_z$), radial direction ($\vec{p}_r$) and in tangential one ($\vec{p}_\theta$). The model assumes that oscillation of the dimer is in the tangential, that is $z-\theta$ plane, which means that $\vec{p}_r \cdot \vec{E}_1 = \vec{p}_r \cdot \vec{E}_2 = 0$ and the relevant moment, used in this paper, is $p = \sqrt{p_z^2 + p_\theta^2} = 391$Debye [17].

The Hamiltonian for a MT can be written as [17]

$$H = \sum_n \left[ \frac{I}{2}\dot{\varphi}_n^2 + \frac{k}{2}(\varphi_{n+1} - \varphi_n)^2 - \frac{A}{2}\theta_n^2 + \frac{B}{4}\theta_n^4 - C\theta_n - pE_1 \cos\varphi_n \right], \quad (2)$$

where $n$ determines a position of the dimer. The first term is kinetic energy, the dot means the first derivative with respect to time and $I$ is a moment of inertia of a single dimer. The second term is the potential energy of the interaction between adjacent dimers belonging to the same PF in the nearest neighbour approximation, where $k$ is the inter-dimer stiffness parameter. The next three terms in Eq. (2) represent a non-symmetric W-potential energy, where $A>0$, $B>0$ and $C>0$ are assumed. This potential determines the directions around which the dimer can oscillate. The very last term in Eq. (2) comes from the fact that the dimer is an electric dipole existing in the field of all other dimers. Of course, $p$ is an electric dipole moment, while $E_1$ is our arbitrary choice. It is assumed that $p>0$ and $E_1>0$.

From Eqs. (1) and (2) and using generalized coordinate $q_n = \varphi_n$ and momentum $p_n = I\dot{\varphi}_n$, as well as Hamilton's equations of motion $\dot{q}_n = \partial H/\partial p_n$, $\dot{p}_n = -\partial H/\partial q_n$, we straightforwardly obtain the following dynamical equation of motion:

$$I\ddot{\varphi}_n = k(\varphi_{n+1} + \varphi_{n-1} - 2\varphi_n) - A_0\varphi_n - C_0\varphi_n^2 - B_0\varphi_n^3 + D_0, \tag{3}$$

where $A_0 = -A + 3B\theta_0^2 + pE_1$, $B_0 = B - pE_1/6$, $C_0 = 3B\theta_0$ and $D_0 = A\theta_0 - B\theta_0^3 + C = 0$ [17].

As was mentioned above, we solve Eq. (3) using the SDA [18]. This mathematical procedure was explained including a lot of details in Ref. [19]. Its mathematical basis is a multiple-scale method or a derivative-expansion method [20,21].

According to the SDA, we assume small oscillations, i.e.

$$\varphi_n = \varepsilon \Phi_n, \quad \varepsilon \ll 1, \tag{4}$$

which changes Eq. (3) into

$$I\ddot{\Phi}_n = k(\Phi_{n+1} + \Phi_{n-1} - 2\Phi_n) - A_0\Phi_n - \varepsilon C_0\Phi_n^2 - \varepsilon^2 B_0\Phi_n^3. \tag{5}$$

A key point in the procedure is that we expect the solution to be a modulated wave, i.e. in the form

$$\Phi_n(t) = F(\xi)e^{i\theta_n} + \varepsilon\left[F_0(\xi) + F_2(\xi)e^{i2\theta_n}\right] + \text{cc} + \text{O}(\varepsilon^2), \tag{6}$$

$$\xi = (\varepsilon nl, \varepsilon t), \quad \theta_n = nql - \omega t, \tag{7}$$

where $F(\xi)$ and $e^{i\theta_n}$ represent an envelope and a carrier components, respectively. The function $e^{i\theta_n}$ obviously includes discreteness, while the envelope will be treated in a continuum limit. As the frequency of the carrier wave is much higher than the frequency of the envelope, we need two time scales, $t$ and $\varepsilon t$, for those two functions. Of course, the same holds for the coordinate scales. In Eqs. (6) and (7), $\omega$ is the optical frequency of the linear approximation, $q = 2\pi/\lambda$ is the wave number, cc stands for complex conjugate terms and $F_0$ is real.

A rather tedious mathematics [17] shows that the functions $F_0$ and $F_2$ can be expressed through $F$, that is $F_0 = \mu |F|^2$ and $F_2 = \delta F^2$, while $F$ is a solution of nonlinear Schrödinger equation (NLSE). The expressions for $\mu$ and $\delta$ are given in Ref. [17]. All this brings about a final result

$$\varphi_n(t) = 2A'\text{sech}\left(\frac{nl - V_e t}{L}\right)\left\{\cos(\Theta nl - \Omega t) + A'\text{sech}\left(\frac{nl - V_e t}{L}\right)\right.$$
$$\left.\times \left[\frac{\mu}{2} + \delta\cos(2(\Theta nl - \Omega t))\right]\right\}, \tag{8}$$

where

$$V_e = V_g + U_e, \qquad U_e = \frac{P}{1-\eta}\left[-q + q\sqrt{1 + \frac{2(1-\eta)}{Pq^2}(\omega - qV_g)}\right]. \tag{9}$$

The parameters $P$ and $Q$ are the dispersion coefficient and coefficient of nonlinearity, respectively. They are given in Ref. [17], as well as the expressions for $A'$, $L$, $\Theta$ and $\Omega$. The second expression in Eq. (9) was obtained based on the idea of a coherent mode (CM) [15,17], assuming that the envelope and carrier wave velocities are equal, that is $V_e = \Omega/\Theta$. Due to the equality of these velocities, the function $\varphi_n(t)$ is the same at any position $n$. This tempting idea has been used for years. However, recent numerical calculations show that this might not always be the case [17]. This interesting problem certainly requires further research. The meaning of the parameter $\eta$ was explained in Ref. [15]. Its allowed interval is

$$0 \leq \eta < 0.5. \tag{10}$$

To plot the function $\varphi_n(t)$ the values of all parameters should be known or estimated. This rather tedious job was performed in Ref. [17]. The function is shown in Fig. 3 for a certain allowed combination of the parameters. This is obviously a localized modulated wave. One can see that, for the chosen values of the parameters, the angle $\varphi$ takes the values from about $-6°$ to $8°$, while the wave covers about 112 dimers. The solitonic speed, corresponding to this example, is $V_e = 445\,\text{m/s}$ [17].

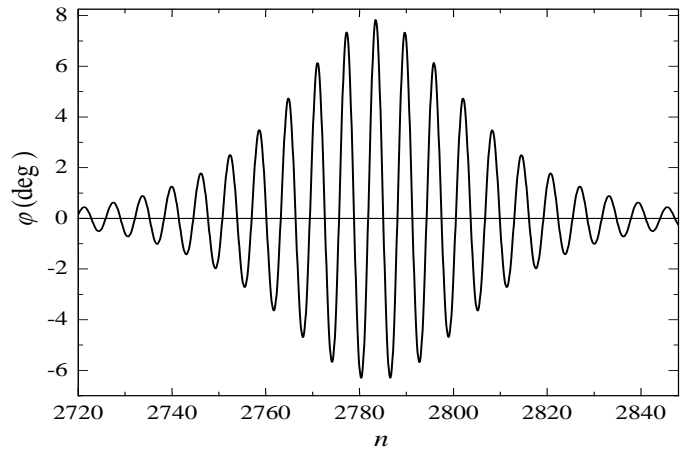

**Fig. 3.** The function $\varphi_n(t)$ as a function of $n$ for $t = 50\text{ns}$, $k = 12\text{eV}$, $q = \pi/7l$ and $\eta = 0.495$. The CM is assumed

As a conclusion, we should point out that the model explained here is a two component one in a sense that the variable $\varphi$ describes the oscillation of the dimer around the direction of the electric field, while $\theta$ determines the orientation of the field. The used SDA assumes a continuum limit [17]. A question if MT is predominantly discrete or continuum system was studied in Ref. [22].

The model explained here is obviously a mechanical one, but this does not mean that MTs are mechanical systems only. There are experiments which indicate electrodynamic activity of variety of cells in the frequency region from kHz to GHz, expecting MTs to be the source of this activity [4,23]. Therefore, MTs are both mechanical and electrical systems and, regarding their modelling, the best that should be done is to work towards more component models taking both characteristics into consideration. One such attempt is the model introduced in Ref. [24].

MTs can also be modelled as nonlinear RLC transmission lines [25-27]. Electrical activities of MTs are very important in fighting some diseases [28]. It is known that MTs can behave as biomolecular transistors capable of amplifying electrical information [29]. This may affect some crucial neuronal computational capabilities, such as memory and consciousness [29,30].

Finally, it might be interesting to mention kinocilium, a component of vestibular hair cells of the inner ear, comprising 10 pairs of MTs [31].

This is a sensory apparatus that receives the environmental signals and transmit them via collectively excited conformational changes in MTs [31]. Of course, this is possible due to the fact that MTs are capable of specific type of wave propagation, as explained above.

## 3. Kinks and bell-type solitons in DNA

There are a lot of models describing complex DNA dynamics [32,33]. The first nonlinear one was introduced in 1980, suggesting that nonlinear effects may focus the vibration energy of DNA into localized soliton-like excitations [34].

In this section, we rely on the well-known helicoidal Peyrard-Bishop (HPB) model for DNA dynamics [19,35]. This is an extended version of the PB model, which does not take helicoidal structure into consideration [36]. It might be important to mention that, in some papers, the HPB model is called Peyrard-Bishop-Dauxois (PBD) model. However, there is a similar model [37,38] called the PBD one and, consequently, it is more convenient to name it the HPB model.

As was explained above, the first step is Hamiltonian, from which we obtain the dynamical equation of motion. In Section 2, we demonstrated the SDA for solving it, the method that has been used for years to study DNA dynamics [19,35]. However, the CA was used recently [39] and this is what we explain in this section.

Fig. 4 shows a segment of DNA chain. Interactions along the strands are strong and the longitudinal displacements are neglected. The relevant ones at the position $n$ are $u_n$ and $v_n$, obviously along the weak hydrogen bonds. Keeping all this in mind, we can write the Hamiltonian as [19,35,39]

$$H = \sum \left\{ \frac{m}{2}(\dot{u}_n^2 + \dot{v}_n^2) + \frac{k}{2}[(u_n - u_{n-1})^2 + (v_n - v_{n-1})^2] \right.$$
$$\left. + \frac{K}{2}[(u_n - v_{n+h})^2 + (u_n - v_{n-h})^2] + D[e^{-a(u_n - v_n)} - 1]^2 \right\}, \quad (11)$$

where $m = 5.4 \cdot 10^{-25}$ kg is the average nucleotide mass, a dot means the first derivative with respect to time, the parameters $k$ and $K$ are coupling constants of the harmonic longitudinal and helicoidal springs, respectively. The first term obviously represents kinetic energy, the second one is the potential energy of the covalent bond, while the third term describes helicoidal interactions. Namely, due to the helicoidal structure, a nucleotide belonging to one strand at the position $n$ comes close to the $n+h$ nucleotide from the other strand. We assume $h=5$ because the helix has a helical pitch of about 10 base pairs per turn [40]. The last term in Eq. (11) is Morse potential energy, describing the weak interaction, where the parameters $D$ and $a$ are the depth and inverse width of the Morse potential well, respectively.

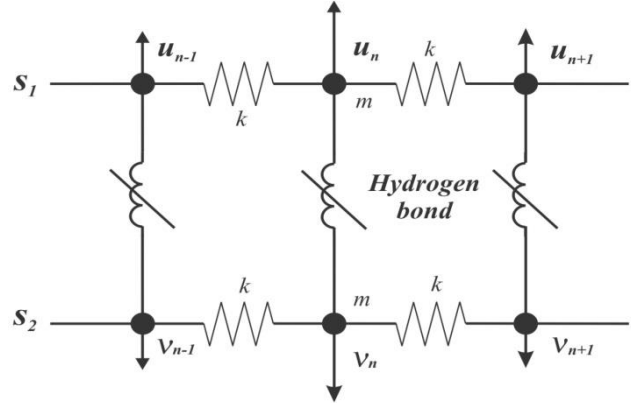

**Fig. 4.** A short portion of DNA molecule

It is convenient to introduce new coordinates $x_n = (u_n + v_n)/\sqrt{2}$ and $y_n = (u_n - v_n)/\sqrt{2}$, representing the in-phase and out-of-phase transversal displacements, respectively. In other words, $x_n(t)$ describes oscillation of the centre of mass of the nucleotide pair, while $y_n(t)$ represents their stretching. From a point of view of DNA activity (breathing, transcription, replication,…) the pair stretching is crucial, which means that we should see DNA molecule as a collection of nucleotide pairs rather than a collection of single nucleotides.

As was explained above, we use Eq. (11) and the Hamilton`s equations of motion, which brings about the following two completely decoupled dynamical equations of motion [19,35,39]

$$m\ddot{x}_n = k(x_{n+1} + x_{n-1} - 2x_n) + K(x_{n+h} + x_{n-h} - 2x_n), \tag{12}$$

$$m\ddot{y}_n = k(y_{n+1} + y_{n-1} - 2y_n) - K(y_{n+h} + y_{n-h} + 2y_n)$$
$$+ 2\sqrt{2}aD(e^{-a\sqrt{2}y_n} - 1)\, e^{-a\sqrt{2}y_n}. \tag{13}$$

The first dynamical equation is a standard linear discrete equation, whose solution is a linear wave (phonon). So, in what follows, we solve Eq. (13) to which we add a viscosity force $-\gamma \dot{y}_n$ on the right side, where $\gamma$ is a viscosity coefficient [9,39,41-44]. We use both the CA $y_n(t) \to y(x,t)$ and appropriate series expansions, which yields to the following nonlinear partial DE

$$m\frac{\partial^2 y}{\partial t^2} - l^2(k - Kh^2)\frac{\partial^2 y}{\partial x^2} + Ay - By^2 + \gamma\frac{\partial y}{\partial t} = 0, \tag{14}$$

where $l = 3.4\,\text{Å}$ is a distance between the two neighbouring nucleotides in the same strand, $A = 4(K + a^2 D)$ and $B = 6\sqrt{2}a^3 D$ [39].

It is well known that, for a given wave equation, a travelling wave $y(\xi)$ is a solution which depends upon $x$ and $t$ through a unified variable $\xi \equiv \kappa x - \omega t$, where $\kappa$ and $\omega$ are constants. This brings about the following ordinary DE [39]

$$\alpha \psi'' - \rho \psi' + \psi - \psi^2 = 0, \qquad \psi' \equiv d\psi/d\xi, \tag{15}$$

where

$$y = (A/B)\psi, \quad \alpha = \frac{m\omega^2 - l^2 \kappa^2 (k - Kh^2)}{A}, \quad \rho = \frac{\gamma\omega}{A}. \tag{16}$$

There are many procedures for solving Eq. (15). Some of them are: standard procedure [9,45], modified extended tanh-function (METHF) method [46-48], method of factorization [49-51], procedure based on Jacobian elliptic functions [52,53], the simplest equation method (SEM) [54-56], modified SEM [57], exponential function procedure [58,59], ($G'/G$)-expansion method [60,61], etc. Except the standard procedure and method of factorization, in all mentioned methods the function $\psi$ is expected to be a series of known functions. However, the series expansion in terms of unknown functions is also possible [62,63]. In this section, we explain METHF method, probably the simplest procedure representing series expansions in terms of known functions. According to this procedure, we look for possible solutions of Eq. (15) in the form

$$\psi = a_0 + \sum_{i=1}^{M}\left(a_i \Phi^i + b_i \Phi^{-i}\right), \tag{17}$$

where the function $\Phi = \Phi(\xi)$ is a solution of the well-known Riccati equation [46-48]

$$\Phi' = b + \Phi^2. \tag{18}$$

The parameters $a_0$, $a_i$, $b_i$ and $b$ are real constants that should be determined, as well as the cut off integer $M$. One can easily show that, for Eq. (15), $M = 2$ [39]. We are looking for the solutions having physical sense and assume

$$\Phi = -\sqrt{-b}\tanh\left(\sqrt{-b}\,\xi\right), \qquad b_i = 0, \tag{19}$$

which is the solution of Eq. (18) for $b < 0$.

According to Eqs. (17) and (18), we obtain expressions for $\psi'$, $\psi''$ and $\psi^2$ and Eq. (15) becomes $A_1\Phi + A_2\Phi^2 + A_3\Phi^3 + A_4\Phi^4 + A_0 = 0$, where $A_i$, $i = 0,...,4$, are coefficients depending on the parameters $b$, $a_0$, $a_1$, $a_2$ and $\alpha$ [39]. Of course, this equation is satisfied if all these coefficients are simultaneously equal to zero, which gives a system of five equations. This system brings about the following two solutions [39]

$$a_0^{(1)} = 1/4,\ \alpha^{(1)} = 6\rho^2/25,\ a_0^{(2)} = 3/4,\ \alpha^{(2)} = -6\rho^2/25,\ a_2^{(i)} = 6\alpha^{(i)}, \tag{20}$$

while the remaining parameters are common for both of them, i.e.

$$a_1 = -\frac{6\rho}{5}, \qquad b = -\frac{25}{144\rho^2} < 0. \tag{21}$$

Finally, according to Eqs. (17), (19), (20) and (21), we easily obtain the solutions we are looking for

$$\psi_1(\xi) = \frac{1}{4}\left(1 + 2\tanh w + \tanh^2 w\right),\ \psi_2(\xi) = \frac{1}{4}\left(3 + 2\tanh w - \tanh^2 w\right), \tag{22}$$

where $w = 5\xi/(12\rho)$. These functions are shown in Fig. 5 for $\rho = 5/12$. Numerically and analytically derived kink profiles nicely fit to each other. The numerical solutions were generated applying the simple Runge-Kutta procedure to Eq. (15), which was firstly transformed into the set of two ordinary first order DEs [56]. One can show that $\psi_1$ describes a supersonic kink, while $\psi_2$ corresponds to subsonic one [39]. This is related to the parameter $\alpha$. Namely, Eq. (16) can be written as

$$\alpha = m\kappa^2\left(V^2 - c^2\right)/A, \qquad c^2 = l^2\left(k - Kh^2\right)/m, \tag{23}$$

where $V = \omega/\kappa$ and $c$ are the solitonic and linear sound velocities, respectively. As $A$ in Eq. (23) is positive, we conclude that the negative $\alpha$ corresponds to the subsonic soliton and vice versa. Some estimations of the velocities $V$ and $c$ can be found in Ref. [39].

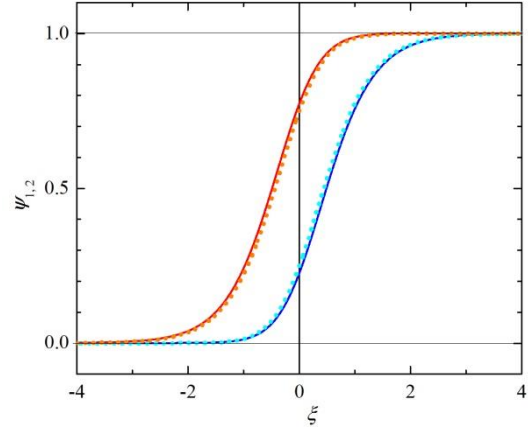

**Fig. 5.** Solutions $\psi_1(\xi)$ (blue) and $\psi_2(\xi)$ (red) for $\rho = 5/12$. The solid and dotted lines correspond to the numerically and analytically derived kinks, respectively

It might be interesting to study the solutions of Eq. (15) when viscosity is neglected. Applying the same procedure as above for $\rho = 0$ we obtain the following two solutions [39]

$$\psi_{10}(\xi) = \frac{1}{2}\left[-1 + 3\tanh^2\left(\sqrt{\frac{3}{2a_2^{(1)}}}\xi\right)\right], \qquad a_2^{(1)} > 0, \qquad (24)$$

$$\psi_{20}(\xi) = \frac{3}{2}\left[1 - \tanh^2\left(\sqrt{-\frac{3}{2a_2^{(2)}}}\xi\right)\right], \qquad a_2^{(2)} < 0. \qquad (25)$$

Obviously, these solutions are expressed through the parameters $a_2^{(1)}$ and $a_2^{(2)}$, and they are shown for $a_2^{(1)} = -a_2^{(2)} = 3/2$ in Fig. 6. These are bell-type solitons. The numerical solutions were generated applying the same procedure as for Fig. 5. The functions $\psi_{10}(\xi)$ and $\psi_{20}(\xi)$ represent the supersonic and subsonic solitons, respectively.

What has been shown so far is that the kinks and bell-type solitons may exist in DNA under certain conditions. A key question is which one really, or, at least, very likely exists. In other words, we should deal with stability of the mentioned solutions. We have performed a series of numerical simulations of Eq. (13) with and without viscosity term [39]. Eqs. (22), (24) and (25) represent initial conditions and the system was checked during 10ps [39]. Fig. 7 shows the function $y_1(\xi)$ corresponding, of course, to $\psi_1(\xi)$ in Fig. 5. The function $y_2(\xi)$ is almost indistinguishable from $y_1(\xi)$, as expected from Fig. 5. Basically, this numerical solution matches analytical one but the obtained kink decreases in time.

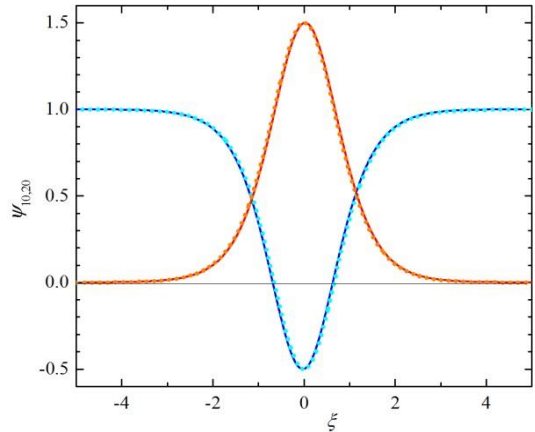

**Fig. 6.** Solutions $\psi_{10}(\xi)$ (blue) and $\psi_{20}(\xi)$ (red) for $\left|a_2^{(i)}\right| = 3/2$, $i = 1, 2$. The solid and dotted lines correspond to the numerically and analytically derived solitons, respectively

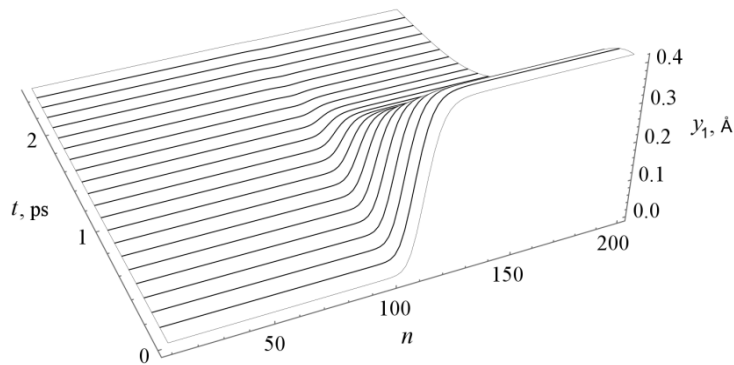

**Fig. 7.** Solution $y_1(\xi)$ for $\gamma = 2.8 \times 10^{-12}$ kg/s

When viscosity is neglected, the solutions (24) and (25) completely change their structures in time. An example is shown in Fig. 8. The solution $y_{20}(\xi)$, as well as $y_{10}(\xi)$, is obviously unstable. This should not bother us because these functions describe non-realistic case when viscosity is neglected. Therefore, if we compare Figs. 7 and 8 we can conclude that the solitons $y_1(\xi)$ and $y_2(\xi)$ are acceptable, while $y_{10}(\xi)$ and $y_{20}(\xi)$ are not. This certainly shows that viscosity is crucial for the wave stability. However, our positive attitude towards Fig. 7 should be discussed. First of all, the soliton decreases in time, which means that it is not stable, at least it is not mathematically stable. However, it exists during a certain period of time and, from biological point of view, a question is if it can perform a required biological task during its lifetime. Let us assume that the kink $y_1$ lives about $1.5\,\text{ps}$, which is suggested by Fig. 7. Its speed was estimated to be $V_2 = 1350\,\text{m/s}$ [39]. During this period of time the kink passes over the distance of about 6 nucleotide pairs. This value can be compared with the experimental value for RNA:DNA hybrid, which is about 8 pairs [64]. These two values match rather well, which means that the kinks $y_1(\xi)$ and $y_2(\xi)$ are biologically acceptable. We are going to return to the RNA:DNA hybrid in the next section.

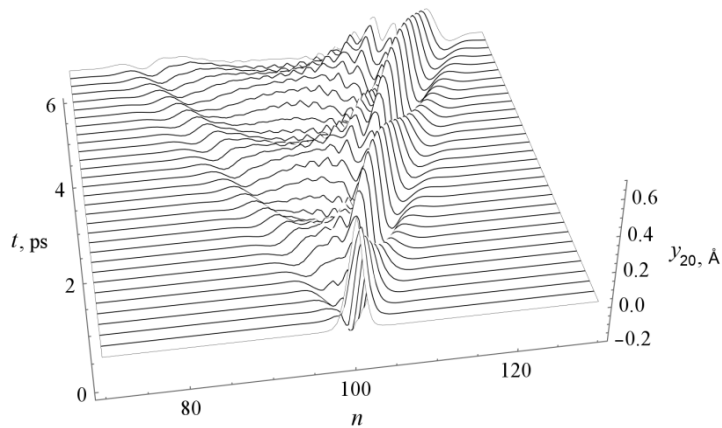

**Fig. 8.** Solution $y_{20}(\xi)$

## 4. Demodulated standing solitary wave and DNA-RNA transcription

In the previous two sections, we studied two biological systems and two mathematical methods. The combinations were MT-breather and DNA-kink. In this section, we study DNA using the SDA, i.e. the combination DNA-breather. Our goal is to study DNA-RNA transcription and it turned out that it may be possible within the idea of existence of the breathers in the chain [65]. We rely on the HPB model again and follow Ref. [65].

It is known that the transcription occurs at the segments of DNA chain that are surrounded by RNA polymerase molecules (RNAP), which is shown in Fig. 9 [66,67]. Let us call these segments transcription segments (TSs). One can see that one of the two DNA strands serves as a template for synthesis of a new RNA strand. It is important to know that the transcription is possible because DNA molecule opens locally at these segments, which implies significantly smaller coupling between base pairs. It was shown that the local opening could be seen as DNA breathing mode with extremely high amplitude [68], which, otherwise, can be conceived as a resonance mode [69].

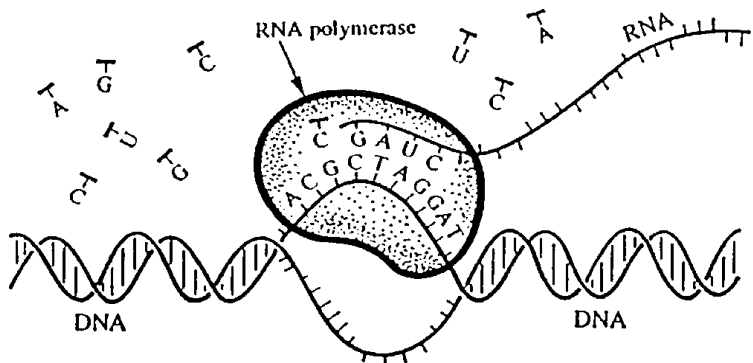

**Fig. 9.** DNA-RNA transcription (Taken from Ref. [66]).

The main goal of this section is to study DNA breathing at TSs in the context of two ideas. We explain why it would be biologically convenient if the soliton were demodulated at TSs. Our second idea is that the soliton becomes a standing one at the TSs. Hence, we can talk of demodulated standing solitary (DSS) mode. We believe that this mode decreases probability for genetic mistakes and yields to successful transcription.

Therefore, we deal with DNA and Eqs. (11) and (13) hold again. We use the SDA explained in Section 2, which means that we assume small oscillations ($y_n = \varepsilon \Phi_n$, $\varepsilon \ll 1$) as well as Eqs. (6) and (7). The final result is the function

$$y_n(t) = 2A\operatorname{sech}\left(\frac{nl - V_e t}{L}\right)\left\{\cos(\Theta nl - \Omega t) + A\operatorname{sech}\left(\frac{nl - V_e t}{L}\right)\right.$$
$$\left.\left[\frac{\mu}{2} + \delta \cos(2(\Theta nl - \Omega t))\right]\right\}, \tag{26}$$

which is, practically, the same as Eq. (8) except that $\varphi_n(t)$ has been replaced by $y_n(t)$. Of course, the expressions for $P$, $Q$, $\mu$ and $\delta$ are different and given in Ref. [65], while Eqs. (9) and (10) hold again.

Let us get back to transcription and study a certain TS. When DNA gets copied into RNA, RNAP attaches itself to one of the two DNA strands, as shown in Fig. 9. This means that RNAP pulls nucleotides out of solution and form RNA according to DNA order of basis. Therefore, we can talk of DNA and RNA nucleotides.

Let us concentrate on one DNA adenine, for example. Normally, it is bonded with DNA thymine belonging to other strand but also interacts with RNA nucleotides, as can be seen from Fig. 9. The final positioning of RNA nucleotides should be a certain copy of the DNA segment, which means that our DNA adenine should attract a certain RNA uracile and repel the remaining RNA nucleotides [65]. This can be efficiently done only if the DNA adenine is far enough from its DNA partner during transcription, which is really the case due to the local opening.

We argued that the local opening is necessary but not sufficient condition for successful transcription [65]. The stretching of DNA, i.e. the distance between the DNA nucleotides belonging to the same pair, is described by Eq. (26). This obviously means that the respective DNA thymine and adenine are far from each other only during short periods of time and the chosen adenine does not have enough time to attract one RNA uracile. The carrier wave is crucial for soliton movement along DNA chain but is redundant when transcription occurs. Also, it makes sense to believe that only the envelope of Eq. (26) corresponds to local opening. All this suggests that the breather should be demodulated when it reaches a TS. This, practically, means that we should get rid of the cosine functions in Eq. (26), which means that the conditions

$$\Theta = 0, \qquad \Omega = 0 \tag{27}$$

should be satisfied at TSs [65]. A crucial question is how demodulation happens at these segments. A simple explanation is that RNAP changes chemical milieu for DNA nucleotides, i.e. the values of relevant parameters, especially $D$ and $a$, which yields to the values accommodating Eq. (27). That DNA surrounding, which is, practically, viscosity, can lead to demodulation was shown in Ref. [70]. This section could be understood as a mathematical analysis of this discovery.

One more idea was suggested recently [65]. Both local opening and demodulation increases time during which the DNA and RNA

nucleotides interact. This is probably not enough but there is one more mechanism to increase this time. Namely, this time is bigger if the soliton velocity is smaller. Hence, biologically convenient soliton is the one which is as slow as possible at the TSs and we have proposed the idea that the soliton wave becomes a standing one at these segments. By the standing wave we assume the one for which the envelope velocity is equal to zero, that is

$$V_e = 0. \tag{28}$$

There have been some suggestions how to experimentally determine the soliton speed, width and even its character [71,72]. They are based on micromanipulation experiments on the single DNA molecule [73-81]. Unfortunately, the expressions (27) and (28) have been neither approved nor disapproved so far. What theoreticians can do is to study if the DSS mode is possible [65]. In particular, we investigate if there exists a certain value of $q$ satisfying Eqs. (27) and (28). We introduce new parameters $x$ and $p$ defined as $K = xk$ and $a^2 D = pk$ [65] and use $h = 5$, as explained earlier. Both $x$ and $p$ should be much less than one because $k$ determines the strong covalent interaction. There are a couple of requirements that should be satisfied, such as $\eta < 0.5$, $P > 0$, $Q > 0$, etc. [65]. For each of them we find intervals for $ql$ satisfying it. For example, we plot the function $P(ql)$ for different values of $x$ and determine the accepted intervals for $ql$. In the end, we compare all these intervals and obtain the final result, which is [65]

$$\left.\begin{array}{l}0.577 < ql < 0.578 \quad \text{and} \\ 0.65 < ql < 0.84 \quad \text{for} \quad x = 1/30, \\ 0.47 < ql < 0.81 \quad \text{for} \quad x = 1/50, \\ \quad ql < 0.77 \quad \text{for} \quad x = 1/80\end{array}\right\}. \tag{29}$$

One can notice an extremely narrow interval for $x = 1/30$. Such intervals do not exist for $x < 1/39.7$ [65]. Also, the lower limit for $ql$ does not exist for $x < 1/62.5$ [65].

Two examples are shown in Fig. 10 for $D = 0.07 \text{eV}$ and $k = 12 \text{N/m}$ [19]. We see the demodulated waves whose amplitudes are $A_1 = 6.1 \overset{\circ}{\text{A}}$ and $A_2 = 1.6 \overset{\circ}{\text{A}}$. The appropriate wave widths are $\Lambda_1 = 8l$ and $\Lambda_2 = 7.8l$, respectively [65]. This means that these waves cover about 8 base pairs, which perfectly matches the experimental value for the DNA:RNA hybrid [64]. Notice close result regarding the kinks in the previous section. The big amplitudes are in agreement with the local opening of the chain. The solitons in Fig. 10 have almost equal widths but their amplitudes vary remarkably. This is so because the amplitude depends on the arbitrary and still unknown parameter $k$. An idea how to experimentally determine $k$ was offered in Ref. [65].

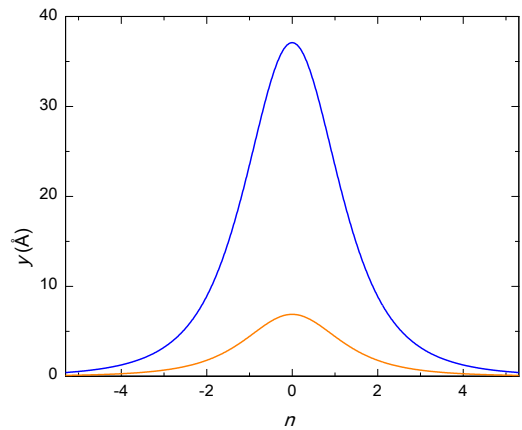

**Fig. 10.** Demodulated solitary wave for $ql = 0.47 \text{rad}$, $x = 1/50$ (blue) and $ql = 0.15 \text{rad}$, $x = 1/80$ (orange)

Therefore, we showed that the values for $ql$, satisfying our postulates explained above, exist. The results are in excellent agreement with the experimental values. A patient reader may have noticed that the big amplitudes are not in agreement with the HPB model, which assumes small amplitudes. This means that the used model predicts the local opening but is not adequate for quantitative analysis. Also, viscosity has been neglected in this section

## 5. Conclusion

Nonlinear dynamics of biological nanosystems is very interesting and developing branch of science. We here studied two of them and explained two mathematical methods. Nonlinearity has been manifested through the solitary waves.

Internal structures of nucleotides and dimers were neglected. As these are relatively big particles, classical physics was used. However, if their internal structures are taken into consideration then quantum mechanics becomes relevant. A common example could be ab initio calculations. Also, if we study charge transfer processes in these systems we should use quantum mechanical approach [82,83].

**References**

[1]   P. Dustin, Microtubules, Springer, Berlin, 1984.

[2]   S. Zdravković, *J. Serb. Chem. Soc.* **82** (2017) 469.

[3]   S. Sahu, S. Ghosh, K. Hirata, D. Fujita and A. Bandyopadhyay, *Appl. Phys. Lett.* **102** (2013) 123701.

[4]   D. Havelka, M. Cifra, O. Kučera, J. Pokorný and J. Vrba, *J. Theor. Biol.* **286** (2011) 31.

[5]   S. Hameroff and R. Penrose, *Phys. Life Rev.* **11** (2014) 39.

[6]   J. E. Schoutens, *J. Biol. Phys.* **31** (2005) 35.

[7]   E. Nogales, M. Whittaker, R. A. Milligan and K. H. Downing, *Cell* **96** (1999) 79.

[8]   P. Drabik, S. Gusarov and A. Kovalenko, *Biophys. J.* **92** (2007) 394.

[9]   M. V. Satarić, J. A. Tuszyński and R. B. Žakula, *Phys. Rev. E* **48** (1993) 589.

[10]  S. Zdravković, L. Kavitha, M. V. Satarić, S. Zeković and J. Petrović, *Chaos Solitons Fract.* **45** (2012) 1378.

[11]  S. Zdravković, S. Zeković, A. N. Bugay and M. V. Satarić, *Appl. Math. Comput.* **285** (2016) 248.

[12]  S. Zdravković and G. Gligorić, *Chaos* **26** (2016) 063101.

[13]  S. Zdravković, A. N. Bugay and A. Yu. Parkhomenko, *Nonlinear Dynam.* **90** (2017) 2841.

[14]  S. Zdravković, M. V. Satarić, A. Maluckov and A. Balaž, *Appl. Math. Comput.* **237** (2014) 227.

[15] S. Zdravković, A. N. Bugay, G. F. Aru and A. Maluckov, *Chaos* **24** (2014) 023139.

[16] S. Zdravković, M. V. Satarić and V. Sivčević, *Nonlinear Dynam.* **92** (2018) 479.

[17] S. Zdravković, S. Zeković and A. N. Bugay, Tangential model of microtubules and semi-discrete approximation, *in preparation*.

[18] M. Remoissenet, *Phys. Rev. B* **33** (1986) 2386.

[19] S. Zdravković, *J. Nonlin Math. Phys.* **18**, Suppl. 2 (2011) 463.

[20] R. K. Dodd, J. C. Eilbeck, J. D. Gibbon and H. C. Morris, Solitons and Nonlinear Wave Equations, Academic Press, Inc., London, 1982.

[21] T. Kawahara, *J. Phys. Soc. Japan* **35** (1973) 1537.

[22] S. Zdravković, A. Maluckov, M. Đekić, S. Kuzmanović and M. V. Satarić, Appl. Math. Comput. **242** (2014) 353.

[23] M. Cifra, J. Pokorný, D. Havelka and O. Kučera, *BioSystems* **100** (2010) 122.

[24] A. N. Bugay, *Nonlin. Phenomena Complex Sys*. **18** (2015) 236.

[25] M. V. Satarić, D. I. Ilić, N. Ralević and J. A. Tuszynski, *Eur. Biophys. J.* **38** (2009) 637.

[26] P. G. Ghomsi, J. T. T. Berinyoh and F. M. M. Kakmeni, *Chaos* **28** (2018) 023106.

[27] F. II Ndzana and A. Mohamadou, *Chaos* **29** (2019) 013116.

[28] F. T. Ndjomatchoua, C. Tchawoua, F. M. M. Kakmeni, B. P. Le Ru and H. E. Z. Tonnang, *Chaos* **26** (2016) 053111.

[29] A. Priel, A. J. Ramos, J. A. Tuszynski and H. F. Cantiello, *Biophys. J.* **90** (2006) 4639.

[30] M. R. Cantero, C. V. Etchegoyen, P. L. Perez, N. Scarinci and H. F. Cantiello, *Sci. Rep.-UK* **8** (2018) 11899.

[31] M. V. Sataric, D. L. Sekulic, B. M. Sataric and S. Zdravković, *Prog. Biophys. Mol. Bio.* **119** (2015) 162.

[32] L. V. Yakushevich, Nonlinear Physics of DNA, Wiley Series in Nonlinear Science, John Wiley, Chichester, 1998.

[33] G. Gaeta, C. Reiss, M. Peyrard and T. Dauxois, *Riv. Nuovo Cimento* **17** (1994) 1.

[34] S. W. Englander, N. R. Kallenbach, A. J. Heeger, J. A. Krumhansl and S. Litwin, *Proc. Natl. Acad. Sci. (USA)* **777** (1980) 7222.

[35] T. Dauxois, *Phys. Lett. A* **159** (1991) 390.

[36] M. Peyrard and A. R. Bishop, *Phys. Rev. Lett.* **62** (1989) 2755.

[37] T. Dauxois, M. Peyrard and A. R. Bishop, *Phys. Rev. E* **47** (1993) R44.

[38] M. Peyrard, *Nonlinearity* **17** (2004) R1.

[39] S. Zdravković, D. Chevizovich, A. N. Bugay and A. Maluckov, *Chaos* **29** (2019) 053118.

[40] T. R. Strick, M. N. Dessinges, G. Charvin, N. H. Dekker, J. F. Allemand, D. Bensimon and V. Croquette, *Rep. Prog. Phys.* **66** (2003) 1.

[41] S. Zdravković and M. V. Satarić, *Chin. Phys. Lett.* **24** No.5 (2007) 1210.

[42] C. B. Tabi, A. Mohamadou and T. C. Kofané, *Chin. Phys. Lett.* **26** (2009) 068703.

[43] J. B. Okaly, A. Mvogo, R. L. Woulaché and T. C. Kofané, *Commun. Nonlinear Sci. Numer. Simulat.* **55** (2018) 183.

[44] V. Vasumathi and M. Daniel, *Phys. Rev. E* **80** (2009) 061904.

[45] A. Gordon, *Physica B* **146** (1987) 373.

[46] E. Fan, *Phys. Lett. A* **277** (2000) 212.

[47] S. A. El-Wakil and M. A. Abdou, *Chaos Solitons Fract.* **31** (2007) 840.

[48] A. H. A. Ali, *Phys. Lett. A* **363** (2007) 420.

[49] O. Cornejo-Perez and H. C. Rosu, *Prog. Theor. Phys.* **114** (2005) 533.

[50] O. Cornejo-Pérez, J. Negro, L. M. Nieto and H. C. Rosu, *Found. Phys.* **36** (2006) 1587.

[51] W. Alka, A. Goyal and C. N. Kumar, *Phys. Lett. A* **375** (2011) 480.

[52] C. Dai and J. Zhang, *Chaos Solitons Fract.* **27** (2006) 1042.

[53] S. Zeković, S. Zdravković, L. Kavitha and A. Muniyappan, *Chin. Phys. B* **23** (2014) 020504.

[54] N. A. Kudryashov, *Phys. Lett. A* **342** (2005) 99.

[55] N. A. Kudryashov, *Chaos Solitons Fract.* **24** (2005) 1217.

[56] S. Zdravković and G. Gligorić, *Chaos* **26** (2016) 063101.

[57] N. A. Kudryashov and N. B. Loguinova, *Appl. Math. Comput.* **205** (2008) 396.

[58] M. N. Alam, M. G. Hafez, M. A. Akbar and H.-O. Roshid, *J. Sci. Res.* **7** (2015) 1.

[59] M. N. Alam and F. B. M. Belgacem, *Mathematics* **4** (2016) 1.

[60] M. N. Alam, M. G. Hafez, F. B. M. Belgacem and M. A. Akbar, *Nonlinear Stud.* **22** (2015) 613.

[61] M. N. Alam, F. B. M. Belgacem and M. A. Akbar, *J. Appl. Math. Phys.* **3** (2015) 1571.

[62] A. J. M. Jawad, M. D. Petković and A. Biswas, *Appl. Math. Comput.* **217** (2010) 869.

[63] S. Zdravković and S. Zeković, *Chin. J. Phys.* **55** (2017) 2400.

[64] J. Gelles and R. Landick, *Cell* **93** (1998) 13.

[65] S. Zdravković, M. V. Satarić, A. Yu. Parkhomenko and A. N. Bugay, *Chaos* **28** (2018) 113103.

[66] C. R. Calladine, H. R. Drew, B. F. Luisi and A. A. Travers, Understanding DNA-The Molecule and How It Works, Elsevier Academic Press, Third Edition, 2004.

[67] T. Lipniacki, *Phys. Rev. E* **60** (1999) 7253.

[68] S. Zdravković and M. V. Satarić, *Europhys. Lett.* **78** (2007) 38004.

[69] S. Zdravković and M. V. Satarić, *Europhys. Lett.* **80** (2007) 38003.

[70] S. Zdravković, M. V. Satarić and Lj. Hadžievski, *Chaos* **20** (2010) 043141.

[71] S. Zdravković and M. V. Satarić, *Phys. Rev. E* **77** (2008) 031906.

[72] S. Zdravković and M. V. Satarić, *Phys. Lett. A* **373** (2009) 4453.

[73] S. B. Smith, L. Finzi and C. Bustamante, *Science* **258** (1992) 1122.

[74] G. U. Lee, L. A. Chrisey and R. J. Colton, *Science* **266** (1994) 771.

[75] S. B. Smith, Y. Cui and C. Bustamante, *Science* **271** (1996) 795.

[76] U. Bockelmann, B. Essevaz-Roulet and F. Heslot, *Phys. Rev. Lett.* **79** (1997) 4489.

[77] J. F. Allemand, D. Bensimon, R. Lavery and V. Croquette, *Proc. Natl. Acad. Sci. USA* **95** (1998) 14152.

[78] H. Clausen-Schaumann, M. Rief, C. Tolksdorf and H. E. Gaub, *Biophys. J.* **78** (2000) 1997.

[79] T. Lionnet, S. Joubaud, R. Lavery, D. Bensimon and V. Croquette, *Phys. Rev. Lett.* **96** (2006) 178102.

[80] E. A. Galburt, S. W. Grill and C. Bustamante, *Methods* **48** (2009) 323.

[81] F. Mosconi, J. F. Allemand, D. Bensimon and V. Croquette, *Phys. Rev. Lett.* **102** (2009) 078301.

[82] D. Čevizović, S. Galović and Z. Ivić, *Phys. Rev. E* **84** (2011) 011920.

[83] D. Čevizović, Z. Ivić, Ž. Pržulj, J. Tekić and D. Kapor, *Chem. Phys.* **426** (2013) 9.